\documentclass[12pt]{article}
\pdfoutput=1 
\usepackage[dvipdfmx]{graphicx}
\usepackage{amsfonts}
\usepackage[mathscr]{eucal}
\usepackage{dcolumn}
\usepackage{bm}
\usepackage[colorlinks=true,linkcolor=blue,citecolor=blue]{hyperref}
\usepackage{color}
\usepackage{amsmath}

\begin{document}


\newcommand{\gtrsim}{ \mathop{}_{\textstyle \sim}^{\textstyle >} }
\newcommand{\lesssim}{ \mathop{}_{\textstyle \sim}^{\textstyle <} }
\newcommand{\vev}[1]{ \left\langle {#1} \right\rangle }
\newcommand{\bra}[1]{ \langle {#1} | }
\newcommand{\ket}[1]{ | {#1} \rangle }
\newcommand{\EV}{ \ {\rm eV} }
\newcommand{\KEV}{ \ {\rm keV} }
\newcommand{\MEV}{\  {\rm MeV} }
\newcommand{\GEV}{\  {\rm GeV} }
\newcommand{\TEV}{\  {\rm TeV} }
\newcommand{\1}{\mbox{1}\hspace{-0.25em}\mbox{l}}
\newcommand{\Red}[1]{{\color{red} {#1}}}

\newcommand{\lmk}{\left(}  
\newcommand{\rmk}{\right)}
\newcommand{\lkk}{\left[}  
\newcommand{\rkk}{\right]}
\newcommand{\lhk}{\left \{ }  
\newcommand{\rhk}{\right \} }
\newcommand{\del}{\partial}  
\newcommand{\la}{\left\langle} 
\newcommand{\ra}{\right\rangle}
\newcommand{\half}{\frac{1}{2}}

\newcommand{\bea}{\begin{array}}
\newcommand{\eea}{\end{array}}
\newcommand{\beq}{\begin{eqnarray}}
\newcommand{\eeq}{\end{eqnarray}}

\newcommand{\dd}{\mathrm{d}}
\newcommand{\Mpl}{M_{\rm Pl}}
\newcommand{\mg}{m_{3/2}}
\newcommand{\abs}[1]{\left\vert {#1} \right\vert}
\newcommand{\mphi}{m_{\phi}}
\newcommand{\Hz}{\ {\rm Hz}}
\newcommand{\for}{\quad \text{for }}
\newcommand{\Min}{\text{Min}}
\newcommand{\Max}{\text{Max}}
\newcommand{\Kahler}{K\"{a}hler }
\newcommand{\cphi}{\varphi}

\begin{titlepage}

\baselineskip 8mm

\begin{flushright}
IPMU 17-0085
\end{flushright}

\begin{center}

\vskip 1.2cm

{\Large\bf Gauged Q-ball Decay Rates into Fermions}

\vskip 1.8cm

{\large 
Jeong-Pyong Hong$^{a,b}$ and Masahiro Kawasaki$^{a,b}$
}

\vskip 0.4cm

{\it$^a$Institute for Cosmic Ray Research, The University of Tokyo,
5-1-5 Kashiwanoha, Kashiwa, Chiba 277-8582, Japan}\\
{\it$^b$Kavli IPMU (WPI), The University of Tokyo,
Kashiwa, Chiba 277-8583, Japan}

\date{\today}
\vspace{2cm}

\begin{abstract}
We derive the decay rate of a gauged Q-ball into fermions, applying the leading semi-classical approximation. We find that more particles come out from the surface of a gauged Q-ball, compared to the case of a global Q-ball, due to the electric repulsion. We show that, however, the decay rate of a gauged Q-ball is bounded from above due to the Pauli blocking at the surface of the Q-ball, just as in the case of a global Q-ball. We also find that there is a further suppression due to the Coulomb potential outside the Q-ball, which we find to play the role of a potential barrier for the fermions coming from the inside the Q-ball. 
\end{abstract}


\end{center}
\end{titlepage}

\baselineskip 6mm


\section{Introduction
\label{sec:introduction}}
It is known that in the minimal supersymmetric standard model~(MSSM), the baryon asymmetry in our universe can be generated by Affleck-Dine mechanism~\cite{ad,add}, which produces a scalar field condensate with baryon number. In many models such as the gauge mediated SUSY breaking models,  the spatial inhomogeneities of this condensate due to quantum fluctuations grow and fragment into non-topological solitons called Q-balls~\cite{ks,kkins,kkins2}, which are defined as spherical solutions in a global $U(1)$ theory which minimize the energy of the system with a fixed $U(1)$ charge~\cite{c}. In this case, the baryon number generated in Affleck-Dine mechanism is confined inside Q-balls, so that the baryon asymmetry in the universe is generated by baryons emitted by the decay of the Q-balls. 

The Q-ball decay into other particles was first studied by Cohen {\it et al.}~\cite{coh}, who considered the Yukawa theory and calculated the neutrino pair production rate by leading semi-classical approximation treating the Q-ball as the classical background scalar field, where the Q-ball configuration was approximated as a step function. More realistic configurations are considered in Refs.~\cite{nur,ym}. In particular, the production rates of quarks and gravitinos from the Q-balls in supersymmetric theories were derived in Ref.~\cite{ym}, which can be used to estimate baryon-to-dark matter ratio in gauge mediated SUSY breaking models, where the gravitino is dark matter.

While the Q-ball is a non-topological soliton resulting from global $U(1)$ symmetry, whose generalization to local $U(1)$ symmetry was also proposed. The lowest-energy configuration with a fixed local $U(1)$ charge is called gauged Q-ball~\cite{gaugedqball}, which consists not only of the scalar field, but also of the $U(1)$ gauge field. The properties of gauged Q-ball solutions have been studied analytically and numerically in the literature~\cite{gaugedqball,5,7,8,9,10,11,12}, but their decay into other particles has not been considered. In this paper, we derive the decay rate of the gauged Q-ball into fermions, applying the leading semi-classical approximation used in Ref.~\cite{coh} to the gauged Q-ball. We assume that the scalar field in the Q-ball couples to fermions by Yukawa interaction. Since the gauged Q-ball can be interpreted as the electrically charged Q-ball, the decay rate into particles with the charge of the same sign is expected to be enhanced, compared to the case of a global Q-ball, and we show that this is indeed the case. We also show that, on the other hand, the decay rate of the gauged Q-ball is upperly bounded, due to the Pauli blocking at the surface of the Q-ball, just as pointed out in the case of the global Q-ball~\cite{coh}, and even more suppressed due to the Coulomb potential outside the Q-ball, since it plays the role of a potential barrier for the fermions coming from the inside, as we will see later. 

The paper is organized as follows. In Sec.~\ref{sec:bp}, we review some basic properties of gauged Q-ball. In Sec.~\ref{sec:dr}, we present the theoretical setup for gauged Q-balls coupled to the massless fermions and calculate the gauged Q-ball decay rate into massless fermions by using the leading semi-classical approximation. Sec.~\ref{sec:conc} is devoted to the conclusions.
\section{Gauged Q-ball}
\label{sec:bp}
We consider a theory of a complex scalar field $\phi$ coupled to a $U(1)$ gauge field $A_\mu$. The Lagrangian density is written as follows. 
\begin{align}
\mathcal{L}&=(D_\mu\phi)^\ast D^\mu\phi-V(\phi)-\frac14F_{\mu\nu}F^{\mu\nu},\\
D_\mu&\equiv\partial_\mu+ieA_\mu,
\end{align}
where $V(\phi)$ is a scalar potential and $F_{\mu\nu}=\del_\mu A_\nu-\del_\nu A_\mu$. We introduce the following ansatz on $\phi$,
\begin{align}
&\phi(x,t)\equiv \phi(r)e^{-i\omega t},
\label{eq:gqpara}
\end{align}
which is the same parametrization as that of a global Q-ball.
For the gauge field, we find spatially symmetric solution with no magnetic field, or no electric current:
\begin{align}
&A_0=A_0(r),\\
&A_i=0.
\end{align}
The equations of motion are then given by
\begin{align}
\label{eq:eomf}
&\frac{d^2\phi}{dr^2}+\frac2r\frac{d\phi}{dr}+\phi h^2-\frac{dV}{d\phi}=0,\\
&\frac{d^2h}{dr^2}+\frac2r\frac{dh}{dr}-e^2\phi^2h=0,
\label{eq:eomgaugedqball}
\end{align}
where we redefined the gauge field to absorb $\omega$ as $h\equiv-\omega+eA_0$. 
We set boundary conditions as 
\begin{align}
\label{eq:bdcga}
&\phi(\infty)=0,~~ \frac{d\phi}{dr}(0)=0,\\
&A_0(\infty)=0,~~\frac{dA_0}{dr}(0)=0,
\end{align}
especially to avoid singularities at $r=0$. 

\begin{figure}[t]
\begin{minipage}{.5\linewidth}
  \includegraphics[width=\linewidth]{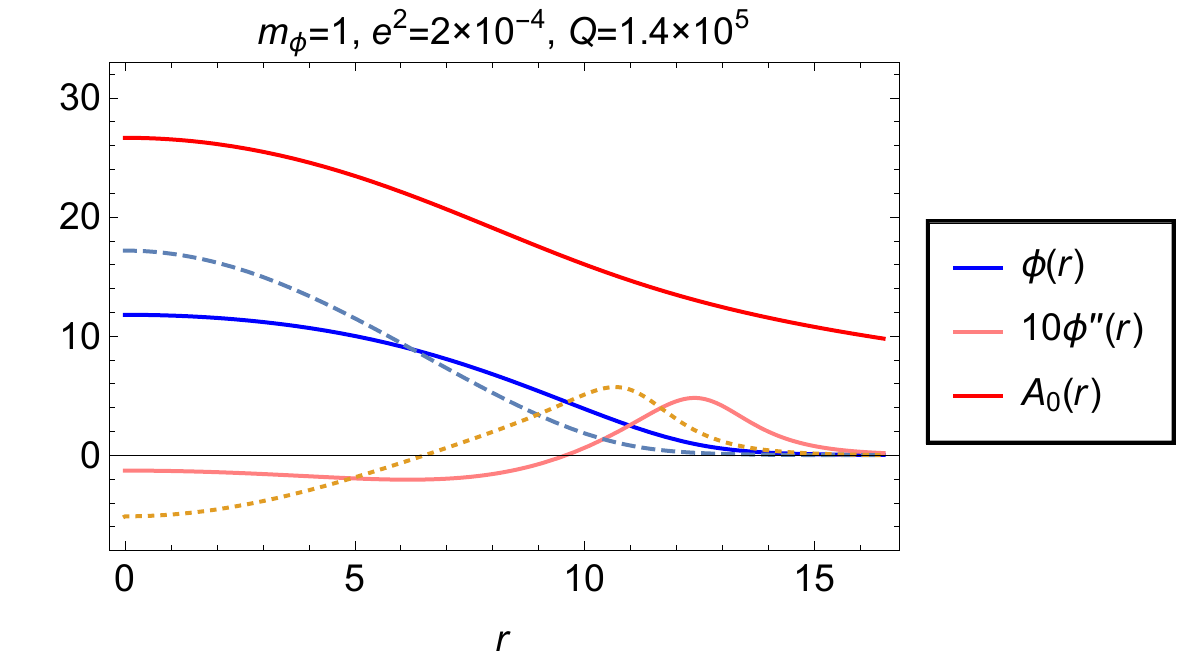}
 \end{minipage}
 \hspace{0cm}
 \begin{minipage}{.5\linewidth}
  \includegraphics[width=\linewidth]{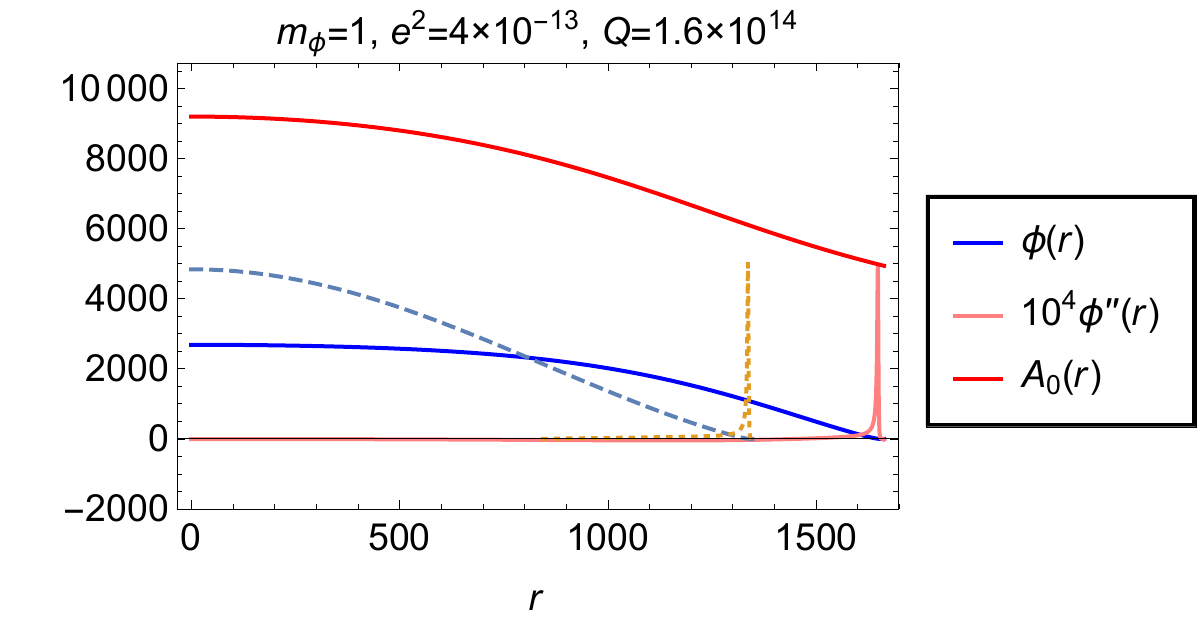}
 \end{minipage}
 \caption{Examples of profile of the gauged Q-ball of gauge mediation type. The dimensionful parameters are in units of $m_\phi$. The dashed line denotes the global Q-ball with the same charge. We see that the profile of $\phi$ is pushed outward due to the electric repulsion. We can also see that $\phi''(r=R)$ becomes singular for a large gauged Q-ball, just as for a global Q-ball, even when the Coulomb potential has a non-negligible effect on the profile.}
\label{fig:gauex}
\end{figure}

As a scalar potential, we choose a logarithmic potential $V(\phi)=m_\phi^4\ln(1+|\phi|^2/m_\phi^2)$, which is motivated by gauge mediation models. For $e=0$, the solution becomes a global Q-ball, which is called gauge mediation type Q-ball~\cite{dv,ls}.
It is known that this type of Q-balls with sufficiently large charge has the following approximate analytic solution.
\begin{align}
\phi(r)=\left\{ 
\begin{array}{ll}
\phi_0\sin\omega r/\omega r,
  & (r\leq R\equiv\pi/\omega)
  \\
  \\
0,  & (r>R)
\end{array}
\right..
\label{eq:sin}
\end{align} 
The angular velocity $\omega$ is equal to $dE/dQ$, which is true for general Q-ball solutions, and has the following charge dependence, 
\begin{align}
\omega=\frac{dE}{dQ}&\propto Q^{-1/4},\label{eq:q14}
\end{align}
which will be useful later.
The second derivative of $\phi$ becomes singular at $r=R$, which for actual Q-balls, becomes a peak of $\phi''(r)$. We define the size of a gauged Q-ball, which is the case $e\neq0$, as the point where $\phi'''(r)=0$ as well, even if the profile is somewhat pushed outward by the electric repulsion, as shown in Fig.~\ref{fig:gauex}. Indeed we can see that $\phi''(r=R)$ becomes singular for a large gauged Q-ball, just as for the global Q-ball, even when the Coulomb potential has a non-negligible effect on the profile. Later we consider the case of large gauged Q-balls when we discuss the saturation of fermion production, where specifying the size of a Q-ball becomes important.

The energy and charge of the gauged Q-ball are given by 
\begin{align}
E&=\int d^3x\left[\frac12(\nabla\phi)^2+\frac1{2}(\nabla A_0)^2+\frac12\phi^2(\omega-eA_0)^2+V(\phi)\right],\\
Q&=\int d^3 x(\omega-eA_0)\phi^2,
\end{align}
and the relation $\omega=dE/dQ$ holds, just as in the case of a global Q-ball, whose proof is given in Ref.~\cite{7}.

\begin{figure}[t]
\begin{minipage}{.47\linewidth}
  \includegraphics[width=\linewidth]{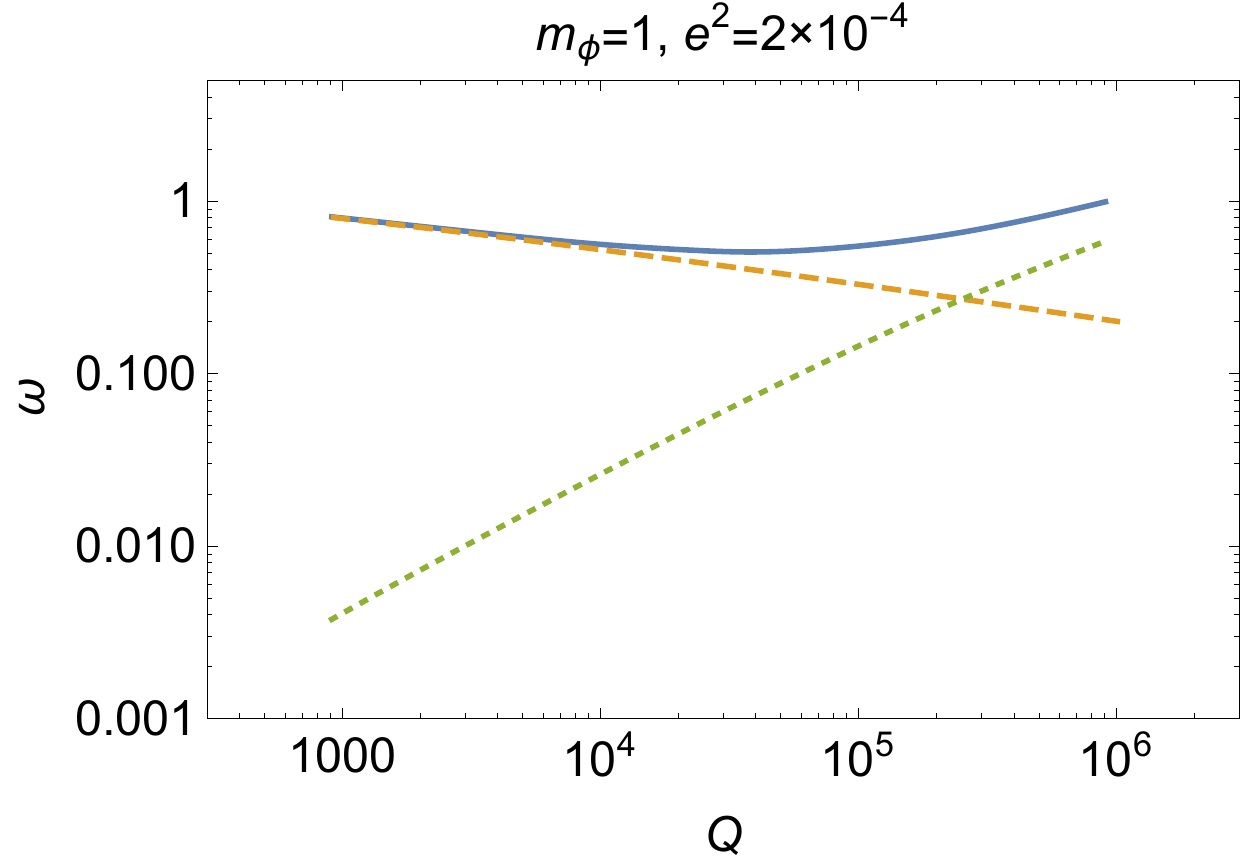}
 \end{minipage}
 \hspace{0.5cm}
 \begin{minipage}{.47\linewidth}
  \includegraphics[width=\linewidth]{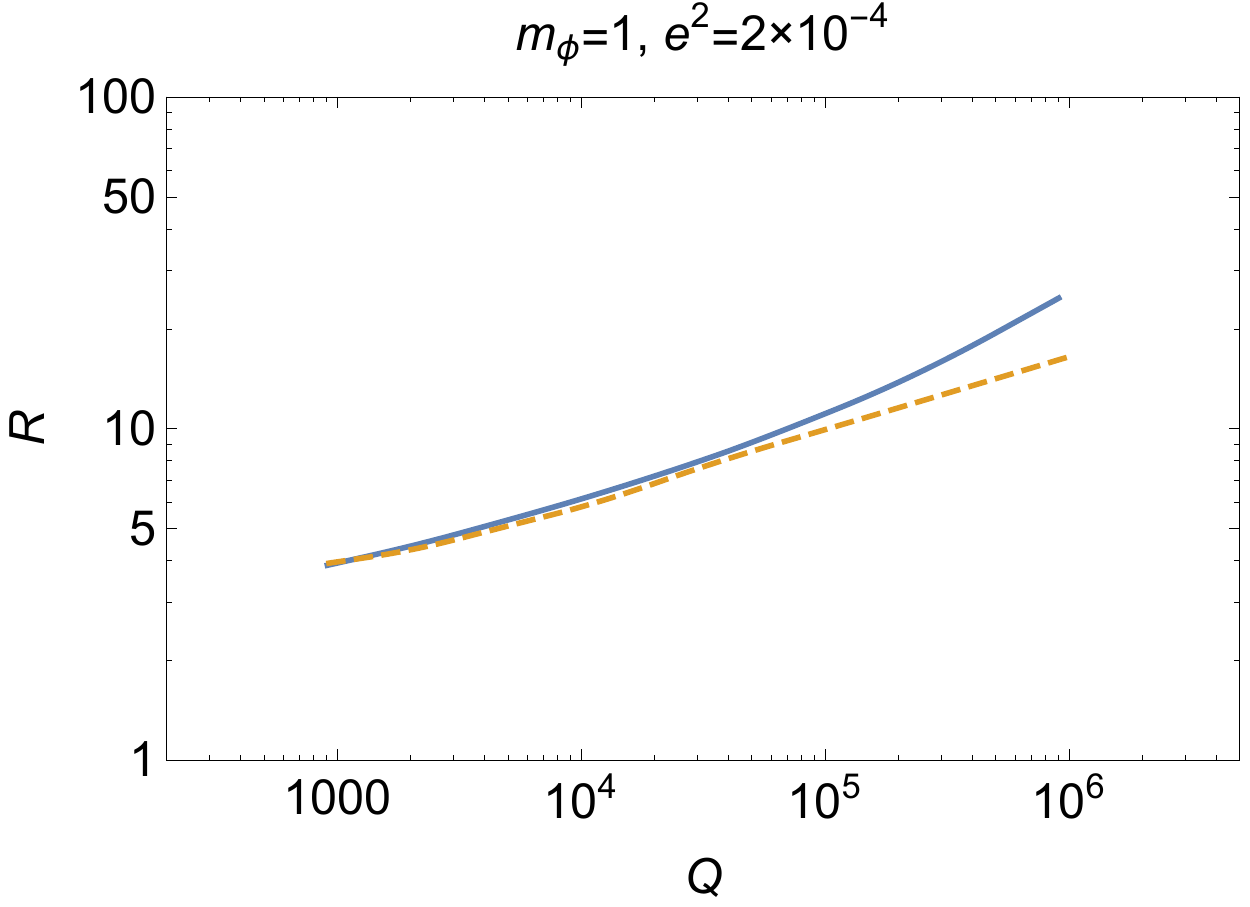}
 \end{minipage}
 \caption{The plots of $\omega=dE/dQ$ and $R$ as functions of $Q$. We plot those for global Q-balls by dashed lines, for comparison. We see that $\omega$ becomes large as the charge grows, which means the Q-ball becomes unstable due to the electric charge, and $R$ becomes large due to the electric repulsion. We also present the Coulomb energy at the surface, $e^2Q/4\pi R$, which is denoted by a dotted line in the left figure.}
\label{fig:omegg}
\end{figure}

The gauged Q-ball becomes unstable as the charge grows due to electric repulsion, which can be seen by the behavior of $\omega=dE/dQ$. We present the plot of $\omega$ as a function of $Q$ in Fig.~\ref{fig:omegg}~(Left). We can see that in contrast to the behavior of $\omega$ for global Q-balls, which is denoted by a dashed line, $\omega$ increases as the charge grows, which means that the Q-ball becomes unstable\footnote{
If the charge grows further, $\omega$ becomes larger than $m_\phi$, which means that the Q-ball becomes unstable against decay into itself, and soon the solutions cease to exist. Here we simply focus on the case $\omega<m_\phi$.}. We also plot Coulomb energy at the surface of Q-ball, $e^2Q/4\pi R$ by a dotted line, whose contribution also becomes large as charge grows. However, we see that the Coulomb energy stays smaller than $\omega$, due to the growth of $\omega$, and also of $R$ by electric repulsion, which we illustrate in Fig.~\ref{fig:omegg}~(Right). 
\section{Gauged Q-ball decay rates into fermions}
\label{sec:dr}
In this section, we derive the decay rate of the gauged Q-ball into fermions, by using the leading semi-classical approximation, where we treat the gauged Q-ball as the classical background. We calculate the production rate of the fermions in the presence of it.  

We consider the following Lagrangian,
\begin{align}
\mathcal{L}_{\text{fermion}}=\chi^\dagger i\bar{\sigma}^\mu(\del_\mu+iq_\chi eA_\mu)\chi+\eta^\dagger i\bar{\sigma}^\mu(\del_\mu+iq_\eta eA_\mu)\eta-(g\phi^\ast\chi\eta+\text{h.c.}),
\end{align}
where $\chi,\eta$ are Weyl fermions, which couple to $\phi$ by Yukawa interaction, and $\bar{\sigma}^\mu=({\bf1},-\sigma^i)$, where $\sigma^i$ are the Pauli matrices. We note that $q_\chi+q_\eta=1$ must be satisfied due to the charge conservation. 
Here we simply set $(q_\chi,q_\eta)=(1,0)$, which assigns the same sign of charge to~$\chi$. 

The equations of motion are written as 
\begin{align}
i\bar{\sigma}^\mu(\del_\mu+iq_\chi eA_\mu)\chi-g\phi\eta^\dagger&=0,\\
i\sigma^\mu(\del_\mu-iq_\eta eA_\mu)\eta^\dagger-g\phi^\ast\chi&=0.
\end{align}
Since $\phi$ is time dependent, the following modes mix with each other.
\begin{align}
\chi&\propto e^{-ik_+t},\\
\eta^\dagger&\propto e^{i(\omega-k_+)t}\equiv e^{ik_-t},
\end{align}
whose equations of motion become
\begin{align}
(k_++iq_\chi eA_0-i\boldsymbol{\sigma}\cdot\nabla)\chi-g\phi(r)\eta^\dagger&=0,\label{eq:em1}\\
(-k_--iq_\eta eA_0+i\boldsymbol{\sigma}\cdot\nabla)\eta^\dagger-g\phi(r)\chi&=0.\label{eq:em2}
\end{align}

First, we consider the case when $\chi,\eta$ are free fields, whose equations of motion are
\begin{align}
(k_+-i\boldsymbol{\sigma}\cdot\nabla)\chi&=0,\\
(-k_-+i\boldsymbol{\sigma}\cdot\nabla)\eta^\dagger&=0.
\end{align}
Then, we can write the following expansion of $\chi,\eta^\dagger$,
\begin{align}
\chi&=\sum_{j,m}\int_0^\infty dk_+\left[a_{\text{in}}(k_+,j,m)e^{-ik_+t}u^{(1)}(-k_+,j,m;{\bf r})\right.\nonumber\\
&\ \ \ \ \ \ \ \ \ \ \ \ \ \ \ \ \ \ \ \ \ \left.+a_{\text{out}}(k_+,j,m)e^{-ik_+t}u^{(2)}(-k_+,j,m;{\bf r})+\text{terms for antiparticle} \right],\\
\eta^\dagger&=\sum_{j,m}\int_0^\infty dk_-\left[(-1)^{m_-}c_{\text{in}}^\dagger(k_-,j,-m)e^{ik_-t}u^{(1)}(-k_-,j,m;{\bf r})\right.\nonumber\\
&\ \ \ \ \ \ \ \ \ \ \ \ \ \ \ \ \ \ \ \ \ \left.+(-1)^{m_-}c_{\text{out}}^\dagger(k_-,j,-m)e^{ik_-t}u^{(2)}(-k_-,j,m;{\bf r})+\text{terms for antiparticle}\right],
\end{align}
using the solution of $(k+i\boldsymbol{\sigma}\cdot\nabla)u^{(i)}=0$, which is defined as 
\begin{align}
u^{(i)}(k,j,m;{\bf r})\equiv\frac{k}{\sqrt{\pi}}\left[h_{l'}^{(i)}(kr)\Phi(j,m,l')+ih_l^{(i)}(kr)\Phi(j,m,l)\right],\ (i=1,2)
\end{align}
where $(l,l')\equiv(j+1/2,j-1/2)$ and $h_l^{(i)}$ denote Spherical Hankel functions. 
$\Phi(j,m,l)$ and $\Phi(j,m,l')$ are the Pauli spinors, which are defined as follows.
\begin{align}
\Phi(j,m,l\equiv j+1/2)&\equiv\left(\begin{array}{ccc}\frac{\sqrt{j-m+1}}{\sqrt{2(j+1)}}Y_l^{m-1/2}\\-\frac{\sqrt{j+m+1}}{\sqrt{2(j+1)}}Y_l^{m+1/2}\\\end{array}\right),\\
\Phi(j,m,l'\equiv j-1/2)&\equiv\left(\begin{array}{ccc}\frac{\sqrt{j+m}}{\sqrt{2j}}Y_{l'}^{m-1/2}\\\frac{\sqrt{j-m}}{\sqrt{2j}}Y_{l'}^{m+1/2}\\\end{array}\right).
\end{align}
We also used 
$\eta^\dagger=i\sigma_2(\eta_\alpha)^\ast$, $i\sigma_2u^{(1,2)}(k,j,m;{\bf r})^\ast=(-1)^{m_+}u^{(2,1)}(k,j,-m;{\bf r})$, where $m_{\pm}\equiv m\pm1/2$. 

One may expect that the fermions outside the gauged Q-ball are described by the solutions above, but the Coulomb field $A_0$, which behaves as $\sim 1/r$ outside the Q-ball, cannot be neglected compared to the fermions, which become spherical waves $\sim e^{ikr}/r$, which are asymptotic forms of the spherical Hankel functions. However, as we derive in the Appendix, $A_0$ only gives an additional phase factor~$e^{iq_{\chi,\eta}e^2Q\log(2kr)}$ to the spherical waves at infinity. Thus, we can still identify incoming and outgoing wave solutions as in the previous paragraph, only corrected by the phase factors. 

The coefficients $a_{\text{out}},c_{\text{out}}^\dagger$ can be written as superpositions of reflecting, and transmuting solutions as follows,
\begin{align}
a_{\text{out}}(k_+,j,m)&=R_\chi(k_+,j)a_{\text{in}}(k_+,j,m)+T_\chi(k_+,j)(-1)^{m_-}c_{\text{in}}^\dagger(k_-,j,-m),\label{eq:nuc}\\
(-1)^{m_-}c_{\text{out}}^\dagger(k_-,j,-m)&=T_\eta(k_-,j)a_{\text{in}}(k_+,j,m)+R_\eta(k_-,j)(-1)^{m_-}c_{\text{in}}^\dagger(k_-,j,-m),
\end{align}
whose coefficients must satisfy the following conditions,
\begin{align}
&|T_\chi(k_+,j)|^2=|T_\eta(k_-,j)|^2,\label{eq:eqce}\\
&|R_\chi(k_+,j)|^2+|T_\chi(k_+,j)|^2=1,\\
&|R_\eta(k_-,j)|^2+|T_\eta(k_-,j)|^2=1,
\end{align}
due to the anticommutation relations of the creation and annihilation operators.

If we define the vacuum $|0_{\text{in}}\rangle$ by $a_{\text{in}}|0_{\text{in}}\rangle=c_{\text{in}}|0_{\text{in}}\rangle=0$ at infinity, we see that the number of outgoing $\chi$ becomes
\begin{align}
\langle0_{\text{in}}|a_{\text{out}}^{\dagger}(k_+,j,m)a_{\text{out}}(k_+',j',m')|0_{\text{in}}\rangle=|T_\chi(k_+,j)|^2\delta(k_+-k_+')\delta_{j,j'}\delta_{m,m'},
\end{align}
using Eq.~(\ref{eq:nuc}), and by summing over the states, the production rate $dQ_i/dt$ is calculated as follows,
\begin{align}
\frac{dQ_i}{dt}=\sum_{j=1/2}\int_0^\omega\frac{dk}{2\pi}(2j+1)\left|T_i(k,j)\right|^2,\ \ \ \ (i=\chi,\eta)\label{eq:dni}
\end{align}
where we averaged the particle number over time using $\delta(0)=T/2\pi$. 
This is the decay rate of the gauged Q-ball into the particle species $i$. Note that $\eta$ with momentum $k_\eta$ must be produced by the same amount as $\chi$ with momentum $\omega-k_\eta$, using Eq.~(\ref{eq:eqce}), which is due to the relation $dE/dQ=\omega$ of the gauged Q-balls.

The coefficients $R_i,T_i$ are determined by matching with the interior solutions, where $\phi,A_0\neq0$. The solutions are written as 
\begin{align}
\chi&=f_\chi(r)\Phi(j,m,l')+ig_\chi(r)\Phi(j,m,l),\\
\eta^\dagger&=f_\eta(r)\Phi(j,m,l')+ig_\eta(r)\Phi(j,m,l),
\end{align}
where, again, we expanded the solutions by the Pauli spinors. We numerically solve for $f_i,g_i$, using Eq.~(\ref{eq:em1}) and (\ref{eq:em2}), under the following boundary conditions.
\begin{align}
f_i'(0)=g_i'(0)=0,
\end{align}
which regularize the solutions at $r=0$.

%

\begin{figure}[t]
\centering
  \includegraphics[width=0.9\linewidth]{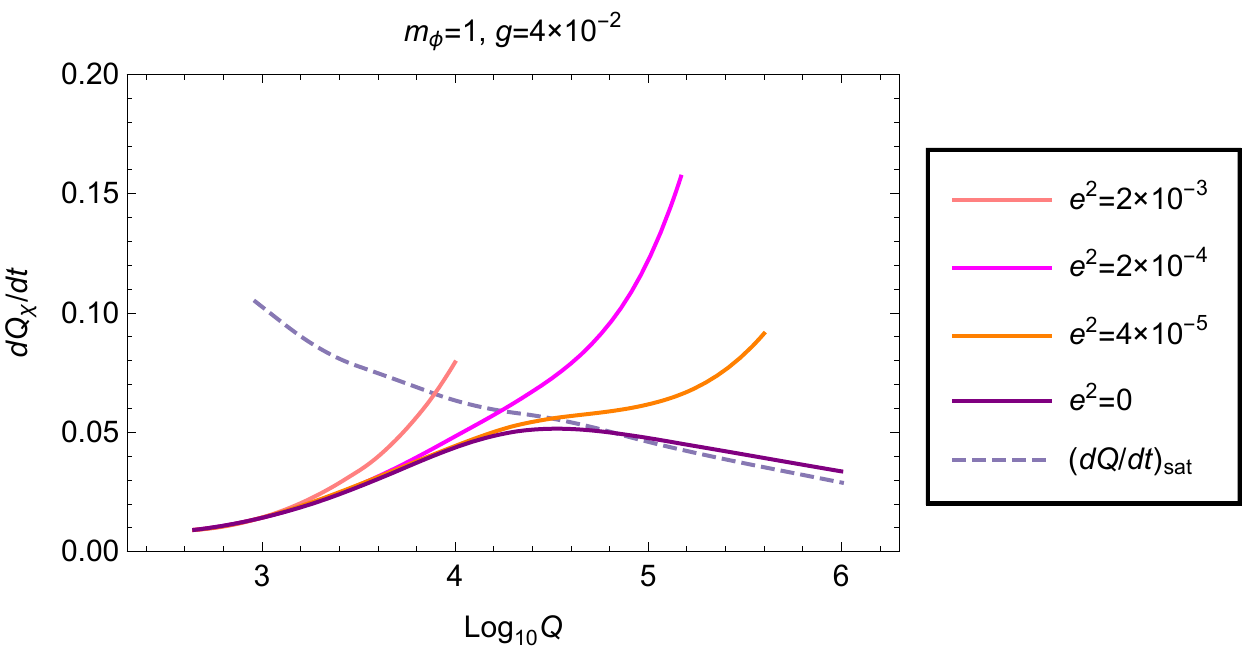}
 \caption{The production rates of fermions from gauged Q-balls. We can see the enhancement due to the electric repulsion. The dashed line indicates the saturated rates for global Q-balls. We can see that for global Q-balls, the production rates saturate as the charge grows, while for gauged Q-balls the saturation is unclear from the figure.}
\label{fig:dn}
\end{figure}

In Fig.~\ref{fig:dn}, we present the results for the production rates. 
Since the gauged Q-ball has electric charge, the decay rate into particles with the charge of the same sign is expected to be enhanced by the electric repulsion, compared to that of the global Q-ball with the same charge. In the figure, we can see that this is indeed the case, where more fermions are
produced for larger gauge coupling $e^2$. 

On the other hand, since $\chi$ is fermion, the flux coming out of the surface of the Q-ball must have an upper bound due to the Pauli blocking. For the global Q-ball, it is obtained by integrating the fully occupied phase space~$(0<k_+<\omega)$, at the surface of the Q-ball~\cite{coh}: 
\begin{align}
\left(\frac{dQ}{dt}\right)_{\text{sat}}\equiv\frac{\omega^3R^2}{24\pi},
\label{eq:sato}
\end{align}
which is called saturated rate. The gauge mediation type global Q-balls have the following properties:
\begin{align}
\omega&\propto Q^{-1/4},\\
R&\simeq\pi/\omega,
\end{align}
Thus, we see that the saturated rate has charge dependence of $Q^{-1/4}$, which is illustrated by a dashed line in the figure. The production rate saturates when the Yukawa interaction becomes strong enough, or when $g\phi_0/\omega\gg1$. Here $\phi_0$ denotes the maximal value of $\phi$. If the Q-ball becomes large, $g\phi_0/\omega$ becomes large so that the interaction effectively becomes strong, which is the reason why the production rate saturates as the charge grows. 

\begin{figure}[t]
\centering
  \includegraphics[width=0.9\linewidth]{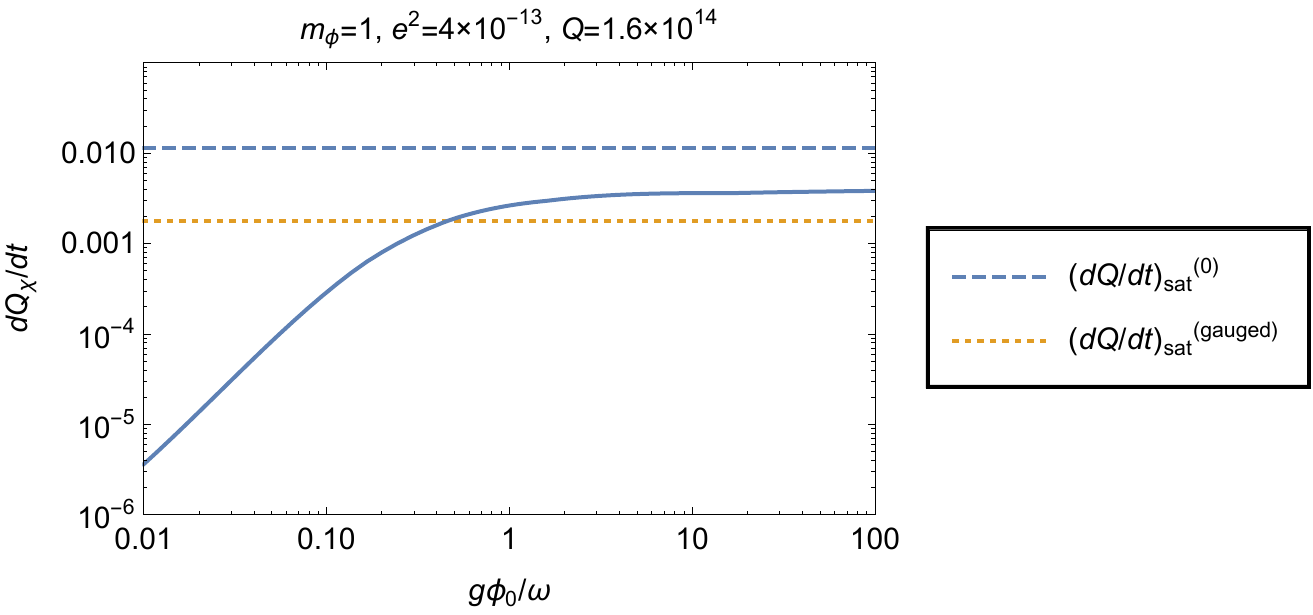}
 \caption{The production rate from a gauged Q-ball as a function of $g\phi_0/\omega$. we see that the production rate saturates for $g\phi_0/\omega\gg1$. We took a gauged Q-ball with large charge, in order to identify the size of the Q-ball as clearly as possible, so that we can compare the production rate to the classically defined saturated rate, which is illustrated by a dotted line. We note that the actual saturated rate is larger than the classical formula, since the fermions with classically forbidden momenta are produced at infinity by quantum tunneling, where Coulomb potential outside effectively becomes potential barrier for fermions coming from inside. We can see that the production rate is suppressed compared to the saturated rate when the Coulomb barrier outside does not exist.}
\label{fig:satto}
\end{figure}

\begin{figure}[t]
\centering
  \includegraphics[width=0.65\linewidth]{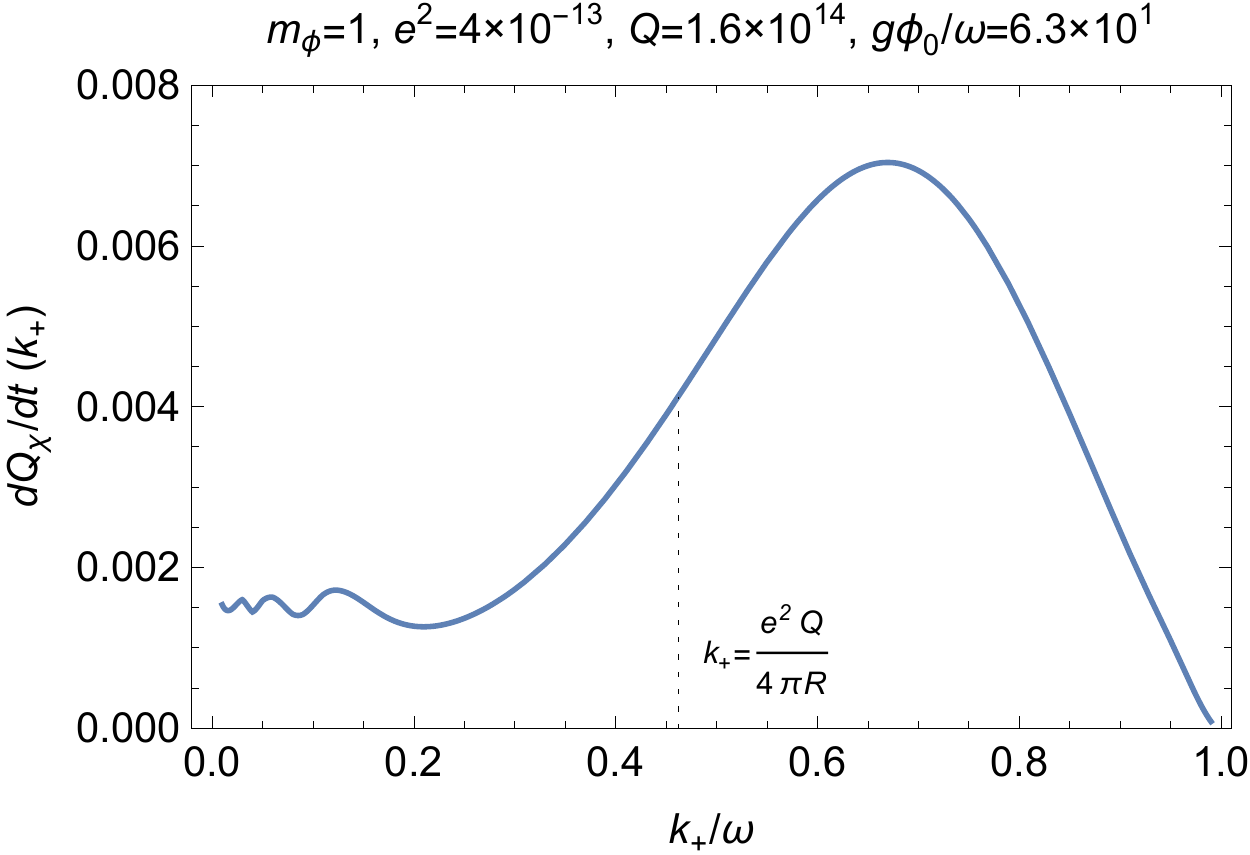}
 \caption{ The production rate as a function of momentum. We can see that the fermions with classically forbidden momenta are produced by quantum effect at infinity.}
\label{fig:tc}
\end{figure}

For gauged Q-balls, however, $\omega$ becomes large as the charge grows, as pointed out in the previous section, hence $g\phi_0/\omega$ does not necessarily become large for a large charge. 
But if we consider a gauged Q-ball with a certain charge and a large Yukawa coupling so that $g\phi_0/\omega\gg1$, we find that the production rate indeed saturates, as shown in Fig.~\ref{fig:satto}. 
We took a gauged Q-ball with large charge, in order to identify the size of the Q-ball as clearly as possible~(see Fig.~\ref{fig:gauex}), so that we can compare the production rate to the saturated rate defined by 
\begin{align}
\left(\frac{dQ}{dt}\right)_{\text{sat}}^{\text{(gauged)}}\equiv\frac{\tilde{\omega}^3R^2}{24\pi},\label{eq:sagau}
\end{align}
where we replaced $\omega$ in Eq.~(\ref{eq:sato}) by $\tilde{\omega}\equiv\omega-e^2Q/4\pi R$, which is the maximal momentum of fermions at the surface of the gauged Q-ball. However, we note that the actual saturated rate is somewhat larger than the one predicted by the above classical formula, whose reason is as follows. Since the classically emitted fermions are ``accelerated", or the momentum is increased by the Coulomb potential outside~($\sim1/r$), the observed fermions must have momentum of $e^2Q/4\pi R<k<\omega$ at infinity. However, as we see in Fig.~\ref{fig:tc}, fermions with momentum of $0<k<e^2Q/4\pi R$ are also observed, which leads to the disagreement in Fig.~\ref{fig:satto}.

The production of fermions with momentum of $0<k<e^2Q/4\pi R$ can be understood as a quantum tunneling effect. If the Yukawa interaction becomes strong, the fermion fields mainly feel $\phi$ inside the Q-ball, which we confirmed numerically as well, and feel the Coulomb potential suddenly at $r=R$. This situation is approximately the same as the case where the fermions produced by $\phi$ come out as a saturated flux with momentum of $0<k<e^2Q/4\pi R$, and bump into the barrier of Coulomb potential at $r=R+\Delta R$~($\Delta R\ll R$), where again, the validity of the approximation is confirmed numerically. This means the production of fermions with momentum of $0<k<e^2Q/4\pi R$ at infinity is due to the tunneling effect, and in particular must be suppressed compared to the saturated rate,
\begin{align}
\left(\frac{dQ}{dt}\right)_{\text{sat}}^{(0)}\equiv\frac{\omega^3R^2}{24\pi},
\label{eq:satos}
\end{align}
when the Coulomb barrier outside does not exist. This can also be confirmed in Fig.~\ref{fig:satto}. 
Thus, we conclude that the decay rate of the gauged Q-ball is bounded from above, due to the Pauli blocking at the surface of the Q-ball, and further suppressed due to the Coulomb potential outside the Q-ball, which effectively becomes a potential barrier for the fermions coming from the inside.

\begin{figure}[t]
\begin{minipage}{.47\linewidth}
  \includegraphics[width=\linewidth]{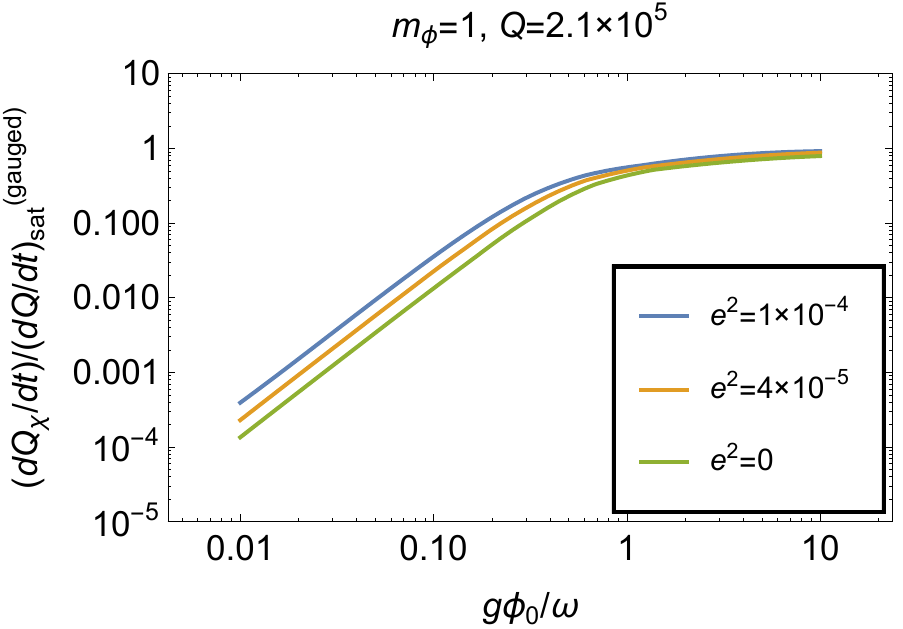}
 \end{minipage}
 \hspace{0.5cm}
 \begin{minipage}{.47\linewidth}
  \includegraphics[width=\linewidth]{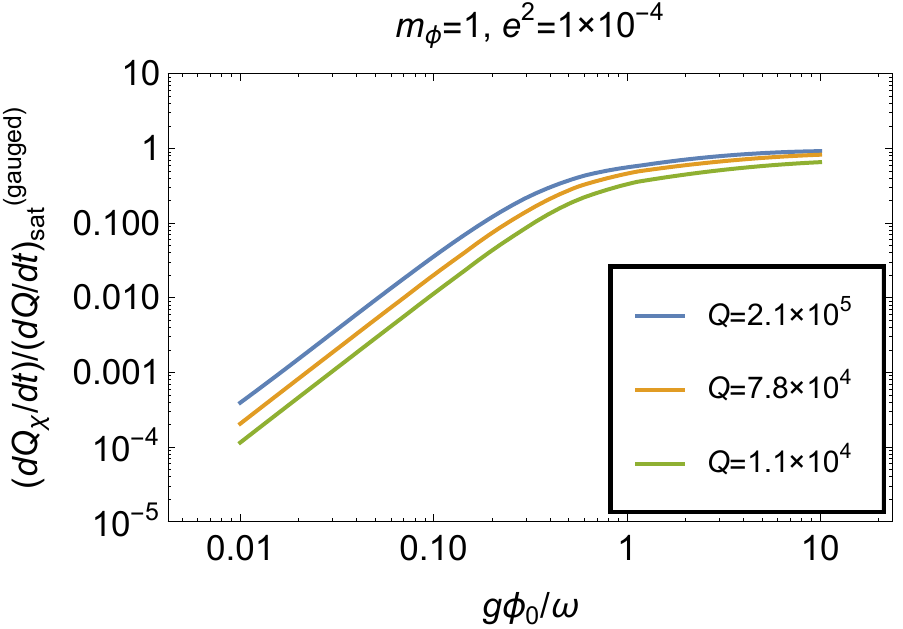}
 \end{minipage}
 \caption{The production rates as a function of $g\phi_0/\omega$, normalized by the classical saturated rates. We find the enhancement of the production rates for $g\phi_0/\omega\ll1$ when $e^2$ or $Q$ becomes large, which can be explained by the analogy to the case of the global Q-ball, where the production rate enhances for a step function-like profile, since the electric repulsion deforms the profile of the gauged Q-ball into a step function-like profile.}
\label{fig:dns}
\end{figure}

Finally, we present the behavior of production rates when $g\phi_0/\omega\ll1$ in Fig.~\ref{fig:dns}. 
Here we consider the gauged Q-balls of weak Coulomb potential with $e^2Q/4\pi R\ll\omega$, and normalize the production rates by the classical saturated rates $\left(dQ/dt\right)_{\text{sat}}^{(\text{gauged})}$, defined by Eq.~(\ref{eq:sagau}). Since the saturated rates have the value between $\left(dQ/dt\right)_{\text{sat}}^{(\text{gauged})}$ and $\left(dQ/dt\right)_{\text{sat}}^{(0)}$, we see that the normalized rate must saturate close to unity for $e^2Q/4\pi R\ll\omega$, which is indeed the case in the figure. Even for such weak Coulomb potential, we note that there are some differences in the production rates when $g\phi_0/\omega\ll1$, depending on the gauge coupling $e^2$, and the charge $Q$. The production rates~(normalized) are enhanced as $e^2$ or $Q$ grows, as shown in the figure. In Ref.~\cite{ym}, it was pointed out that for the global Q-ball, the production rate~(normalized) is enhanced for a step function-like profile. The similar explanation can be valid for the gauged Q-ball as well, since the electric repulsion pushes the charge toward the surface, which makes the profile like a step function. 
\section{Conclusions and discussion}
\label{sec:conc}
In this paper, we derived the decay rate of the gauged Q-ball into fermions, using the semi-classical method in Ref.~\cite{coh}. We assume that the scalar field that forms the gauged Q-ball couples to fermions by Yukawa interaction. 

Since the gauged Q-ball is electrically charged, the decay rate into particles with the charge of the same sign is expected to be enhanced, compared to the case of the global Q-ball. We found that indeed more particles come out from the surface of the gauged Q-ball, compared to the case of the global Q-ball, due to the electric repulsion. 

For global Q-balls, it is known that there is an upper bound on the flux of fermions coming out of the surface of the Q-ball, due to the Pauli blocking, which is called saturated rate. We found that the production rates from each gauged Q-ball also saturate when the Yukawa interaction becomes strong, just as in the case of the global Q-ball. However, the saturated rate is somewhat larger than the one predicted by the classical formula, which is obtained by integrating the fully occupied phase space~$(0<k_+<\omega-e^2Q/4\pi R)$, at the surface of the Q-ball. The disagreement arises since the fermions with classically forbidden momenta are produced by the quantum effect. We found that the production can be interpreted as a tunneling effect, where the fermions, which are mainly produced by $\phi$ due to the strong Yukawa interaction, come out as a saturated flux, and immediately bump into the Coulomb barrier and tunnel through it. The production must be suppressed compared to the saturated rate when the Coulomb barrier outside does not exist, which is also confirmed. 

We also found the enhancement of the production rates~(normalized) for $g\phi_0/\omega\ll1$ when $e^2$ or $Q$ becomes large, which can be explained by the analogy to the case of the global Q-ball, where the production rate enhances for a step function-like profile, since the electric repulsion deforms the profile of the gauged Q-ball into a step function-like profile.

In our previous works, we considered the elecrically charged Q-ball dark matter scenarios~\cite{j1,j2,j3}. In this scenario, the Q-balls formed after the Affleck-Dine mechanism can become electrically charged if the flat direction consists of baryonic and leptonic components and only the leptonic component decays off while the baryonic component is stable. This is possible if the energy of outgoing particle is smaller than the baryon mass, but larger than the lepton mass. Then, we also implicitly assumed that the decay of the leptonic component is sufficiently fast, so that the electrically charged Q-balls are formed in the early universe. From the result of this paper, we can gain some insight on that matter. While we discussed small Q-balls with charge of $10^3$ to $10^6$, for the convenience in the numerical calculations, the Q-balls in the cosmological context, are usually very large, with charge of $10^{20}$ to $10^{30}$. Thus, it is likely that $g\phi_0/\omega\gg 1$ unless the Yukawa coupling $g$ is extremely small, which means that the production rates of leptons are saturated. It was also pointed out in Ref.~\cite{j1} that the electric charge of the Q-ball can grow only until $Q\sim O(100)$ due to the Schwinger effect, etc., and especially the size and the maximal momentum of outgoing particle at the surface are nearly the same as the case of the global Q-ball. Thus, the saturated rate almost does not change from that for the global Q-ball. In all, the decay rate of the leptonic component is approximately written by the saturated rate for the global Q-ball, which is typically known to be of the order of GeV, thus we can conclude that the decay is sufficiently fast, and our previous assumption was reasonable.  

\vspace{1cm}
\section*{Acknowledgments}
J.H. would like to thank Masaki Yamada for helpful comments. This work is supported by MEXT KAKENHI Grant Number 15H05889 (M.K.) and JSPS KAKENHI Grant Number 17K05434 (M.K.). The work is also supported by World Premier International Research Center Initiative (WPI Initiative), MEXT, Japan. 
\vspace{1cm}
\begin{appendix}
\section{Fermions outside the gauged Q-ball}\label{apdix}
It is known that the equations of motion for fermions in the presence of the Coulomb potential of the form~$\sim1/r$ have analytic solutions. Thus, we can give analytic forms for the solution outside the gauged Q-ball. In this appendix, we present the solutions and their asymptotic behaviors, where we especially show that the solutions become spherical waves with some additional phase factors. 

The radial parts of Eq.~(\ref{eq:em1}), which are the equations of motion for $\chi$, are given by 
\begin{align}
\left(k_+-q_\chi \frac{e^2Q}{r}\right)f_\chi+\left(\frac{\del}{\del r}+\frac{3/2+j}{r}\right)g_\chi=0,\label{eq:emap1}\\
\left(k_+-q_\chi \frac{e^2Q}{r}\right)g_\chi-\left(\frac{\del}{\del r}+\frac{1/2-j}{r}\right)f_\chi=0,\label{eq:emap2}
\end{align}
using $\boldsymbol{\sigma}\cdot\nabla\Phi(j,m,j\pm1/2)=\Phi(j,m,j\mp1/2)\left(\del/\del r+(1\pm(j+1/2))/r\right)$, and the solutions are written as the following forms.
\begin{align}
f_\chi(r)&=\frac{e^{ik_+r}}{r}\biggl[r^{s_0}C_+\biggl[{}_1F_1(s_0+iq_\chi e^2Q,2s_0+1;-2ik_+r)\biggr.\biggr.\nonumber\\
&+\left.\frac{s_0+iq_\chi e^2Q}{j+1/2} {}_1F_1(s_0+1+iq_\chi e^2Q,2s_0+1;-2ik_+r)\right]\nonumber\\
&+r^{-s_0}C_-\biggl[{}_1F_1(-s_0+iq_\chi e^2Q,-2s_0+1;-2ik_+r)\biggr.\nonumber\\
&+\left.\left.\frac{-s_0+iq_\chi e^2Q}{j+1/2} {}_1F_1(-s_0+1+iq_\chi e^2Q,-2s_0+1;-2ik_+r)\right]\right], 
\\
g_\chi(r)&=i\frac{e^{ik_+r}}{r}\biggl[r^{s_0}C_+\biggl[{}_1F_1(s_0+iq_\chi e^2Q,2s_0+1;-2ik_+r)\biggr.\biggr.\nonumber\\
&+\left.\frac{s_0+iq_\chi e^2Q}{j+1/2} {}_1F_1(s_0+1+iq_\chi e^2Q,2s_0+1;-2ik_+r)\right]\nonumber\\
&-r^{-s_0}C_-\biggl[{}_1F_1(-s_0+iq_\chi e^2Q,-2s_0+1;-2ik_+r)\biggr.\nonumber\\
&+\left.\left.\frac{-s_0+iq_\chi e^2Q}{j+1/2} {}_1F_1(-s_0+1+iq_\chi e^2Q,-2s_0+1;-2ik_+r)\right]\right], 
\end{align}
where $s_0=\sqrt{(j+1/2)^2-(q_\chi e^2Q)^2}$, and we used the confluent hypergeometric function, which is defined as follows.
\begin{align}
{}_1F_1(a,b;z)\equiv\sum_{k=0}^\infty\frac{a(a+1)\cdots(a+k-1)}{b(b+1)\cdots(b+k-1)}\frac{z^k}{k!}
\end{align}

Using the following asymptotic form of the confluent hypergeometric function,
\begin{align}
{}_1F_1(a,b;z)\sim\Gamma(b)\left(\frac{e^zz^{a-b}}{\Gamma(a)}+\frac{(-1)^{-a}z^{-a}}{\Gamma(b-a)}\right),\,\,\,\,\, |z|\gg1
\end{align}
we find that the solutions behave as
\begin{align}
f_\chi&\sim \left[C_+\frac{s_0+iq_\chi e^2Q}{j+1/2}\left((-1)^{s_0}\frac{(-i)^{-s_0+iq_\chi e^2Q}\Gamma(2s_0+1)}{\Gamma(s_0+1+iq_\chi e^2Q)}\right)\right.\nonumber\\
&\left.\,\,\,\,\,+C_-\frac{-s_0+iq_\chi e^2Q}{j+1/2}\left((-1)^{-s_0}\frac{(-i)^{s_0+iq_\chi e^2Q}\Gamma(-2s_0+1)}{\Gamma(-s_0+1+iq_\chi e^2Q)}\right)\right]\times\frac{e^{-ik_+r+iq_{\chi}e^2Q\log(2kr)}}{r}\nonumber\\
&\,\,\,\,\,+\left[C_+\left((-1)^{s_0}\frac{i^{-s_0-iq_\chi e^2Q}\Gamma(2s_0+1)}{\Gamma(s_0+1-iq_\chi e^2Q)}\right)\right.\nonumber\\
&\left.\,\,\,\,\,\,\,\,\,\,\,\,\,\,\,\,\,\,\,\,\,\,\,\,\,\,\,\,\,\,\,\,\,\,\,\,\,\,\,\,\,\,\,\,\,\,\,\,\,\,+C_-\left((-1)^{-s_0}\frac{i^{s_0-iq_\chi e^2Q}\Gamma(-2s_0+1)}{\Gamma(-s_0+1-iq_\chi e^2Q)}\right)\right]\times\frac{e^{ik_+r-iq_{\chi}e^2Q\log(2kr)}}{r},\\
g_\chi&\sim \left[C_+\frac{s_0+iq_\chi e^2Q}{j+1/2}\left((-1)^{s_0}\frac{(-i)^{-s_0+iq_\chi e^2Q}\Gamma(2s_0+1)}{\Gamma(s_0+1+iq_\chi e^2Q)}\right)\right.\nonumber\\
&\left.\,\,\,\,\,+C_-\frac{-s_0+iq_\chi e^2Q}{j+1/2}\left((-1)^{-s_0}\frac{(-i)^{s_0+iq_\chi e^2Q}\Gamma(-2s_0+1)}{\Gamma(-s_0+1+iq_\chi e^2Q)}\right)\right]\times(+i)\frac{e^{-ik_+r+iq_{\chi}e^2Q\log(2kr)}}{r}\nonumber\\
&\,\,\,\,\,+\left[C_+\left((-1)^{s_0}\frac{i^{-s_0-iq_\chi e^2Q}\Gamma(2s_0+1)}{\Gamma(s_0+1-iq_\chi e^2Q)}\right)\right.\nonumber\\
&\left.\,\,\,\,\,\,\,\,\,\,\,\,\,\,\,\,\,\,\,\,\,\,\,\,\,\,\,\,\,\,\,\,\,\,\,\,\,\,\,\,\,\,\,\,\,\,\,\,\,\,+C_-\left((-1)^{-s_0}\frac{i^{s_0-iq_\chi e^2Q}\Gamma(-2s_0+1)}{\Gamma(-s_0+1-iq_\chi e^2Q)}\right)\right]\times(-i)\frac{e^{ik_+r-iq_{\chi}e^2Q\log(2kr)}}{r},
\end{align}
at infinity, consisting of incoming and outgoing waves with additional phase factors,~$e^{\pm iq_{\chi}e^2Q\log(2kr)}$.

\end{appendix}

\vspace{1cm}




\begin{thebibliography}{90}
\bibitem{ad} I. Affleck and M. Dine, Nucl. Phys. B {\bf249}, 361 (1985)
\bibitem{add} M. Dine, L. Randall, and S. D. Thomas, Nucl. Phys. B {\bf458}, 291 (1996).
\bibitem{ks} A. Kusenko and M. Shaposhnikov, Phys. Lett. B {\bf418}, 46 (1998).
\bibitem{kkins} S. Kasuya and M. Kawasaki, Phys. Rev. D {\bf61}, 041301 (2000).
\bibitem{kkins2} S. Kasuya and M. Kawasaki, Phys. Rev. D {\bf62}, 023512 (2000).
\bibitem{c} S. Coleman, Nucl. Phys. B {\bf262}, 263 (1985).
\bibitem{coh} A. G. Cohen, S. R. Coleman, H. Georgi and A. Manohar, Nucl. Phys. B {\bf272}, 301 (1986).
\bibitem{nur} T. Multamaki and I. Vilja, Nucl. Phys. B {\bf574}, 130 (2000).
\bibitem{ym} M. Kawasaki and M. Yamada, Phys. Rev. D {\bf87}, 023517 (2013).
\bibitem{gaugedqball} K. Lee, J. A. Stein-Schabes, R. Watkins and L. M. Widrow, Phys. Rev. D {\bf39}, 1665 (1989).
\bibitem{5} V. Benci and D. Fortunato, J. Math. Phys. {\bf52}, 093701 (2011).
\bibitem{7} I. E. Gulamov, E. Y. Nugaev and M. N. Smolyakov, Phys. Rev. D {\bf89}, 085006 (2014).
\bibitem{8} C. H. Lee and S. U. Yoon, Mod. Phys. Lett. A {\bf06}, 1479 (1991).
\bibitem{9} H. Arodz and J. Lis, Phys. Rev. D {\bf79}, 045002 (1991).
\bibitem{10} V. Dzhunushaliev and K. G. Zloshchastiev, Central Eur. J. Phys. {\bf11}, 325 (2013).
\bibitem{11} T. Tamaki and N. Sakai, Phys. Rev. D {\bf90}, 085022 (2014).
\bibitem{12} Y. Brihaye, V. Diemer and B. Hartmann, Phys. Rev. D {\bf89}, 084048 (2014).
\bibitem{dv} G. Dvali, A. Kusenko, and M. Shaposhnikov, Phys. Lett. B {\bf417}, 99 (1998).
\bibitem{ls} M. Laine and M. Shaposhnikov, Nucl. Phys. B {\bf532}, 376 (1998).
\bibitem{j1} J. Hong, M. Kawasaki, and M. Yamada, Phys. Rev. D {\bf92}, 063521 (2015).
\bibitem{j2} J. Hong, M. Kawasaki, and M. Yamada, JCAP {\bf1608}, 053 (2016). 
\bibitem{j3} J. Hong and M. Kawasaki, arXiv:1702.00889 [hep-ph].
\end{thebibliography}
\end{document}